\documentclass[twocolumn,showpacs,aps,pre,amssymb,amsmath]{revtex4}

\usepackage{graphicx}
\usepackage{longtable}

\begin{document}

\title{Network structure of chaotic patterns}
\author{Kapilanjan Krishan}
\affiliation{Department of Physics and Astronomy, University of
California - Irvine, Irvine, California 92697-4575}
\date{\today}

\begin{abstract}
We reduce complex stripped patterns to a basic topological network of edges and vertices to define defects and measure their influence on the pattern. We present statistics on the spatial and temporal distribution of defects within the state of spiral defect chaos state in experiments on Rayleigh B\'enard convection. These measure the role of boundary influence on the dynamics, and suggest an exponential distribution for the length of edges in the pattern. We also indicate a systematic method to study hierarchies of defect interactions based on the network structure.
\end{abstract}

\maketitle

Topological variations represent basic changes in the connectivity within the geometric structure of a system. The mechanisms leading to topological change are often associated with physically significant events underlying the dynamics. The reduction of patterns to their discrete topological features is therefore useful in building low dimensional phenomenological models for the dynamics. In some systems, a transition between different geometric structures occurs at spatially and temporally localized regions. For example, the patterns exhibited in Rayleigh-B\'enard convection are observed to transition from regular structures of parallel rolls or target patterns to complex time-dependent chaos through the formation of localized defects~\cite{c89}. These defects mark regions of merger and pinch-off of the stripped patterns exhibited resulting in changes in the topology~\cite{emb98,ct95}. Understanding the dynamics of these defects is therefore important to understanding the irregular and chaotic patterns formed. 

Defects highlight regions where symmetries of a pattern are broken. The identification of a defect therefore often relies on measures of symmetry. In regular convective patterns, defects have been identified by marking the variation in the phase of an underlying parallel striped pattern~\cite{kid04,ss02,ch93}. Spatio-temporal chaotic systems often exhibit intricate patterns with interwoven structure. The absence of symmetries in these patterns makes it difficult to define and identify defects within the structure. In this paper, we present a technique to reduce many conventionally observed complex striped patterns to a discrete set of vertices and edges representing a graph network. This allows fluid systems to be studied using many of the techniques developed in studying complex networks.

\begin{figure}[h!bt]
\includegraphics[width=4.1cm]{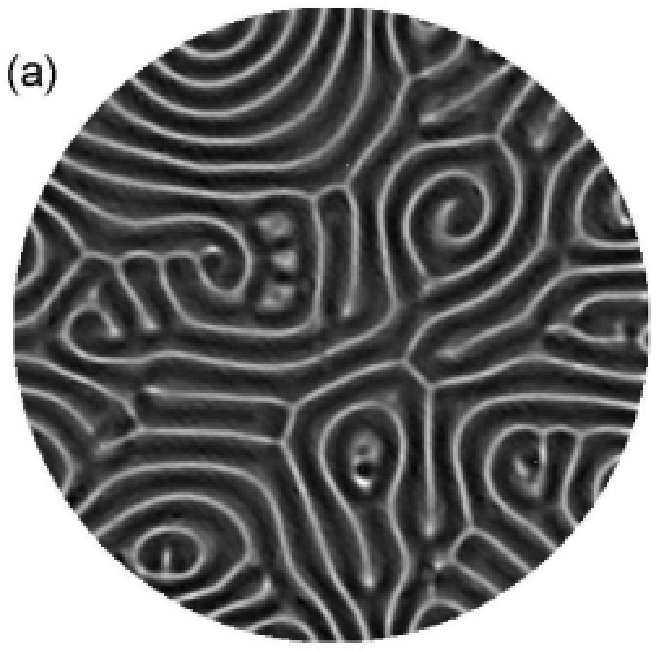}\\
\includegraphics[width=4.1cm]{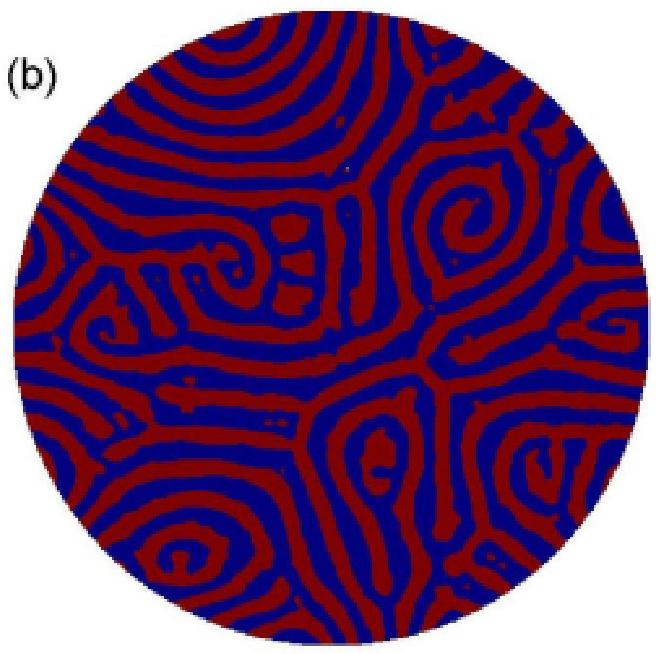}\\
\includegraphics[width=4.1cm]{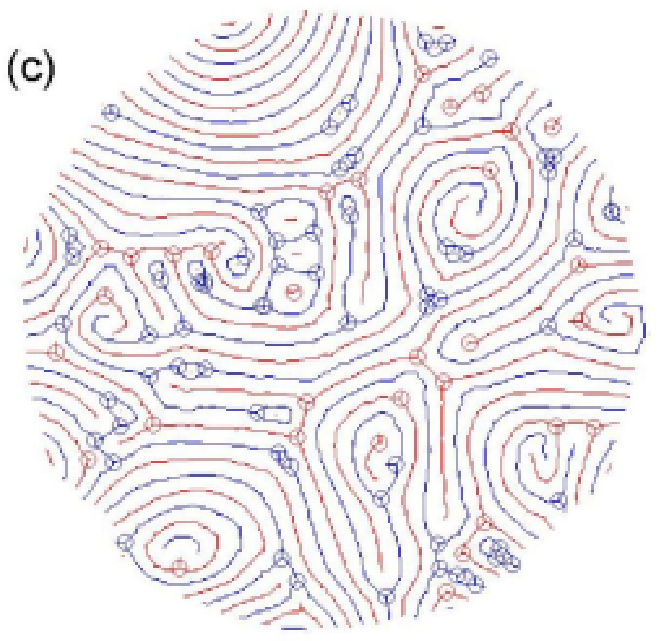}
\caption{The above sequence of images illustrates the two main steps of the image analysis. In (a) a unfiltered image of the convection rolls during spiral defect chaos is shown. This image is filtered and thresholded to yield a binary image with two components of equal area corresponding to bright and dark regions as seen in (b). Bright and dark regions are independently reduced to one dimensional curves, while maintaining the topology of the original pattern. The final skeletonized image is shown in (c). The vertices in the bright and dark regions have been marked by circles.} 
\label{skel}
\end{figure}

The patterns of fluid flow during Rayleigh-B\'enard convection represent among the most thoroughly studied patterns in physics~\cite{c81,ch93}. In this system, a thin layer of fluid(with depth $d$) is held in a vertical thermal gradient(with temperature difference $\Delta T$). At a critical temperature difference, $\Delta T_c$, buoyant forces overcome thermal and viscous dissipation to result in a convective instability. The temperature difference across the cell is parameterized by the non-dimensional reduced Rayleigh number, $\epsilon = (\Delta T - \Delta T_c)/\Delta T_c)$. Ideally, the velocity of the fluid is in the form of a set of parallel rolls. A two dimensional projection of this structure along the horizontal direction yields a parallel stripped pattern. The general, convection rolls form stripes that may be simply or multiply connected regions. A transition from a stationary and regular parallel striped pattern to a time dependent chaotic state known as spiral defect chaos has been observed with an increase in thermal driving~\cite{mbcdg93}.

Our experiments use gaseous CO$_{\rm 2}$ pressurized at $30{\rm atm}$ as the convective fluid confined within a circular convection cell of diameter $3.8 {\rm cm}$ and $d=0.069 {\rm cm}$. We study the patterns exhibited during spiral defect chaos at $\epsilon = 1.1$ to demonstrate the techniques presented in this paper. This state is known as a classic example of spatio-temporal chaos. The temperature difference between hot upflows and cooler downflows are also associated with a variation in refractive index of the gas. These patterns are observed as bright and dark regions in shadowgraphic images~\cite{bbmthca96}. In the experiments, the images are captured using a digital CCD camera with a resolution of 650 pixels $\times$ 515 pixels. Each gray-scale image obtained is Fourier filtered to remove optical distortions inherent in the measurement process (pixel noise within the CCD array for example) and cropped to the circular region of the convection cell. The filtered images are thresholded to their median value of intensity to produce a binary image. This threshold reduces the image to two components with equal areas representing bright and dark regions in the original image(see figure~\ref{skel}b). 

We process each component of the binary image separately. This is done by eroding the pixels at the boundary, except when such erosion would result in connected regions breaking apart. Repeated iteration of the procedure on the components of the binary images reduces their structure to be one pixel thick. Some spurious structures also result with the formation of numerous spurs. These spurs result from a variation in the thickness/wavelength of the convective stripped patterns. These artifacts are removed by further eroding terminating pixels of the structure. This erosion is done to correspond to approximately half the wavelength of the stripped pattern. The final result of the procedure is illustrated in figure~\ref{skel} where the raw experimental image(a), the thresholded image(b) as well as the skeletonized image(c) are indicated. Further details about this morphological procedure is found in~\cite{lls92}.

A limitation inherent to the technique described is that the spatial resolution of the resulting skeletonized features is one pixel. This limitation arises because the skeletonization technique applies to binary images. Sub-pixel spatial resolution may be obtained by weighting the localized regions around desired features by intensity variations in the original gray-scale images. The results reported here are robust within the current limits on spatial resolution.

The image analysis procedure outlined provides an algorithm to reduce two dimensional stripes to one dimensional lines. This skeletonized description maintains the basic topological structure of the pattern by preserving the Euler number. By construction, the one dimensional curves representing bright and dark regions do not intersect each other, however, they often self-intersect. The intersections form vertices along the one dimensional curves and are marked by circles in Figure~\ref{skel}c. These vertices represent regions of defect formation in the pattern. The statistics of defects may therefore be measured through those of the vertices. The shadowgraph images are thus rendered into a network of vertices(defects) and edges(convective rolls) between two defects. Lines without defects also occur in the form of loops or with termination. For the purposes of the current study, we refer to all line segments as edges.

The occurrence of only three fold vertices within the pattern places a severe constraint on the structure of the network. Similar constraints are known in other two-dimensional systems such as foam where local surface tension forces dictate the equilibrium structures or in regular hexagonal packing. In the present case, this tri-fold vertex structure arises from the hydrodynamic stability of the structure~\cite{wpb04}.

\begin{figure}[htb!]
\includegraphics[width=8.6cm]{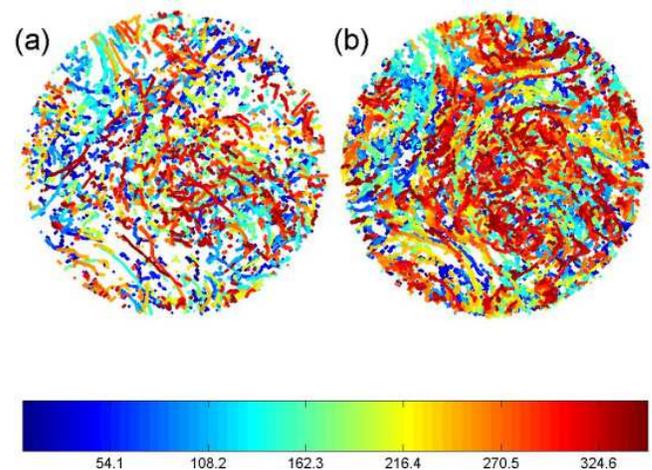}
\caption{The above scatter plot indicates the positions of the defects at different times for bright(a) and dark(b) convection rolls. The colormap indicates the time at which the defects were detected scaled by the vertical diffusion time.} \label{defect_density}
\end{figure}

\begin{figure}[htb!]
\includegraphics[width=8.6cm]{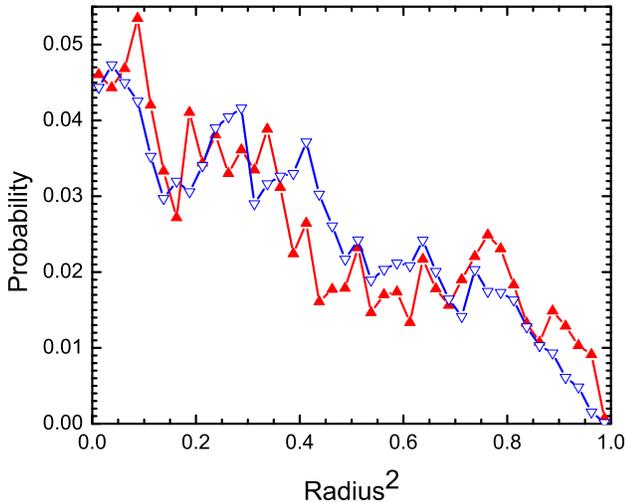}
\caption{The radial distribution of the defect density is shown above. The x-axis is normalized by the radius of the convection cell. The solid and hollow markers represent bright and dark regions of the image.} \label{Rprob}
\end{figure}

During the flow, the spatial locations of the defects change with time. Figure~\ref{defect_density} indicates the variation in position of defects observed within the convection cell. Often, the defects are seen to glide along the convection rolls. The streaks observed in the scatter plot in Figure~\ref{defect_density} are a result of such motion. We also observe that the number of defects is higher for dark regions. This may be because the flow is non-Boussinesq at the parameter values used. The non-Boussinesq parameter (defined in~\cite{b67}) has a value of approximately $0.54$ close to the onset of convection indicating a break in the Boussinesq symmetry. While the number of defects shows a break in the Boussinesq symmetry, the probability distribution of the defects is symmetric as indicated in Figure~\ref{Rprob}.

It is difficult to measure the influence of the boundaries in driving the formation of defects. The radial distribution of defect positions is shown in Figure~\ref{Rprob}. The square of the radius is used along the x-axis as the circular area considered varies quadratically with the radius. The figure shows that the density of defects increases away from the boundaries. Further we calculate that the mean radial velocity of the defects vanishes independent of their radial position, suggesting that the flux of defects moving radially inward is balanced by those moving radially outward. The probability distribution in Figure~\ref{Rprob} is therefore stationary. The stationarity of the distribution implies there is no radially localized source/sink for defect creation/annihilation. We also note that at the aspect ratio studied, the radial distribution of defects does not attain a spatially uniform value. It would be interesting to study the aspect ratio at which a region with spatially uniform defect density is obtained. This region would reflect the defect dynamics inherent to the flow during spiral defect chaos independent of the boundaries.

\begin{figure}[htb!]
\includegraphics[width=8.6cm]{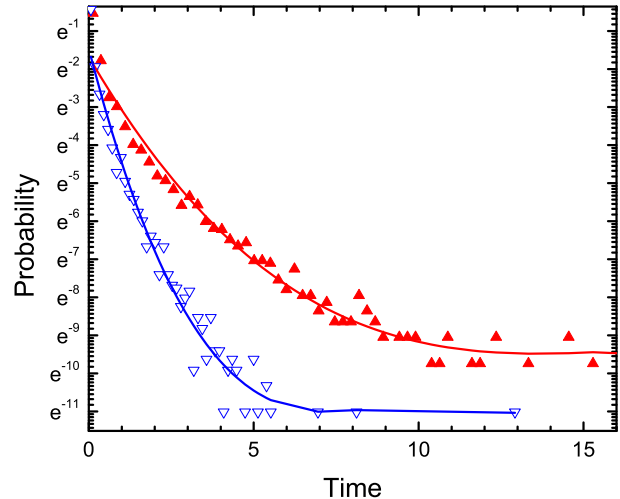}
\caption{The above plot indicates the amount of time each single defect was tracked in units of the vertical thermal diffusion time. The horizontal axis is scaled by the vertical diffusion time. The solid lines are fits to logarithm of the probability using third order polynomials. The coefficients for ($x^3$,$x^2$,$x^1$,$x^0$) are (0.002,0.1,-1.48,-1.68) for the solid markers(bright regions) and (-0.014,0.39,-3.38,-1.46) for the hollow markers(dark regions) respectively.} \label{lifetime}
\end{figure}

In the state of spiral defect chaos, the motion of the convection rolls result in defects being continually created and annihilated. The vertices in the network constructed are created either through the formation of a new edge along an existing edge, or by the intersection of two separate edges. A vertex may be annihilated through the contraction of one of the three edges defining it. When two edges are shared between two vertices, the contraction of the shared edges results in the annihilation of a pair of vertices/defects. Similarly, progressively higher order interactions may be defined, for example, with a vertex triplet, with each vertex sharing a pair of edges with to other vertices(as in a triangle) and so on. The different processes of edge creation and annihilation occur over a range of timescales. We use a standard Particle Image Velocimetry program to identify and track the motion of individual defects during the flow~\cite{piv}.

The lifetime of a defect varies inversely with the number of defects as indicated in figure~\ref{lifetime}. The temporal resolution of image capture in the experiments sets the lower bound of the lifetime measurements. We find that the logarithmic distribution of lifetimes is well fit by a cubic polynomial. The coefficient for the linear term for this fit is large in comparison to those of the quadratic and cubic term which represent small corrections. Also, the number of defects with the same lifetime is larger for bright in comparison to dark stripes in the images. 

The formation of defects in Rayleigh-B\'enard convection has been associated with instability mechanisms about the straight roll state. In our system, the skew-varicose instability and cross-roll instability primarily drive variations in the pattern at short and long wavelengths respectively~\cite{ch93}. The vertex description of a defect does not directly provide information on these instability mechanisms, however, it might be possible to observe differences in the network structure around the neighborhood of defect formation to suggest the type of instability mechanism involved locally.

\begin{figure}[htb!]
\includegraphics[width=8.6cm]{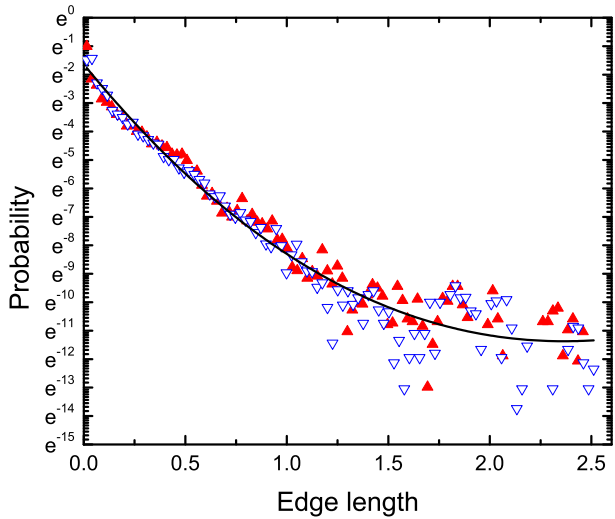}
\caption{The distribution of the edge lengths in the network structure is displayed in the above figure for bright(solid markers) and dark regions(hollow markers) of the convection pattern. A third order polynomial fit(solid line) to the data yields the coefficients to the cubic, quadratic, linear and constant term as (-0.1,2.18,-8.75,-1.64) respectively. The dominant linear term suggests that the edge lengths follow a primarily exponential distribution. The lengths of the edges have been scaled by the diameter of the convection cell.} 
\label{edge_length}
\end{figure}

Each edge in the network is either interspersed with vertices, terminates at the boundary/bulk or very rarely forms a loop with no vertices. The edges in the graph network represent these various states of the convection rolls. The distribution of lengths of the edges is shown in figure~\ref{edge_length}. We find that the number of edges of a given length show a dominantly exponential decay with increasing length of the edges. The deviation from this behavior for edges with large lengths may be due to the finite diameter of the convection cell. The curvature of the convection rolls allow for edges of length exceeding the diameter of the cell. Since the width of bright and dark stripped patterns are similar, and their areas are equal, this also implies that the total length of the edges are equal. Figure~\ref{edge_length} indicates that the sectioning of lengths through defect formation or termination for bright and dark regions are equivalent.

There have been many extensive investigations of defects, their identification and their role in fluid flow. Some of the interesting studies and advances have been cited within the literature referred to in this paper. Of particular interest are recent studies that report measures similar to those extracted in this paper from defect trajectories~\cite{hrdb04}. Geometric diagnostics of chaotic patterns resulting from defect dynamics too have been reported~\cite{rm06}. In addition, the study of patterns through different topological characterizations has been of interest. Extracting rigorous and dimension independent features from pixelated images however can be daunting. The use of computational presents one of the simplest and most easily accessible topological characterization of patterns~\cite{gmk04}. 

Our present study describes a technique to obtain a graph network for images corresponding to experimentally obtained patterns of chaotic fluid flows. These patterns are similar to those obtained from a number of other canonical systems exhibiting spatio-temporal chaos such as reaction-diffusion systems, granular systems, electro-convection in liquid crystals etc. (See~\cite{ch93} for a review). Applying techniques described here lay a common formalism to describe these systems well into the chaotic regime. In the study presented, the patterns are reduced to graphs representing two components of the flow. The vertices of the graph are identified as defects within the pattern. A number of statistics describing the distribution of defects and their influence on the underlying network structure have been described. We have measured the lifetime of a defect and the dispersion of convection roll lengths. We have also quantified the influence of the boundaries on the spatial distribution of the defect dynamics.


\begin{thebibliography}{16}
\expandafter\ifx\csname natexlab\endcsname\relax\def\natexlab#1{#1}\fi
\expandafter\ifx\csname bibnamefont\endcsname\relax
  \def\bibnamefont#1{#1}\fi
\expandafter\ifx\csname bibfnamefont\endcsname\relax
  \def\bibfnamefont#1{#1}\fi
\expandafter\ifx\csname citenamefont\endcsname\relax
  \def\citenamefont#1{#1}\fi
\expandafter\ifx\csname url\endcsname\relax
  \def\url#1{\texttt{#1}}\fi
\expandafter\ifx\csname urlprefix\endcsname\relax\def\urlprefix{URL }\fi
\providecommand{\bibinfo}[2]{#2}
\providecommand{\eprint}[2][]{\url{#2}}

\bibitem[{\citenamefont{Croquette}(1989)}]{c89}
\bibinfo{author}{\bibfnamefont{V.}~\bibnamefont{Croquette}},
  \bibinfo{journal}{Contemporary Physics} \textbf{\bibinfo{volume}{30}},
  \bibinfo{pages}{113} (\bibinfo{year}{1989}).

\bibitem[{\citenamefont{Egolf et~al.}(1998)\citenamefont{Egolf, Melnikov, and
  Bodenschatz}}]{emb98}
\bibinfo{author}{\bibfnamefont{D.~A.} \bibnamefont{Egolf}},
  \bibinfo{author}{\bibfnamefont{I.~V.} \bibnamefont{Melnikov}},
  \bibnamefont{and}
  \bibinfo{author}{\bibfnamefont{E.}~\bibnamefont{Bodenschatz}},
  \bibinfo{journal}{Phys. Rev. Lett.} \textbf{\bibinfo{volume}{80}},
  \bibinfo{pages}{3228} (\bibinfo{year}{1998}).

\bibitem[{\citenamefont{Cross and Tu}(1995)}]{ct95}
\bibinfo{author}{\bibfnamefont{M.~C.} \bibnamefont{Cross}} \bibnamefont{and}
  \bibinfo{author}{\bibfnamefont{Y.}~\bibnamefont{Tu}}, \bibinfo{journal}{Phys.
  Rev. Lett.} \textbf{\bibinfo{volume}{75}}, \bibinfo{pages}{834}
  (\bibinfo{year}{1995}).

\bibitem[{\citenamefont{Kamaga et~al.}(2004)\citenamefont{Kamaga, Ibrahim, and
  Dennin}}]{kid04}
\bibinfo{author}{\bibfnamefont{C.}~\bibnamefont{Kamaga}},
  \bibinfo{author}{\bibfnamefont{F.}~\bibnamefont{Ibrahim}}, \bibnamefont{and}
  \bibinfo{author}{\bibfnamefont{M.}~\bibnamefont{Dennin}},
  \bibinfo{journal}{Physical Review E (Statistical, Nonlinear, and Soft Matter
  Physics)} \textbf{\bibinfo{volume}{69}}, \bibinfo{eid}{066213}
  (pages~\bibinfo{numpages}{6}) (\bibinfo{year}{2004}),
  \urlprefix\url{http://link.aps.org/abstract/PRE/v69/e066213}.

\bibitem[{\citenamefont{Semwogerere and Schatz}(2002)}]{ss02}
\bibinfo{author}{\bibfnamefont{D.}~\bibnamefont{Semwogerere}} \bibnamefont{and}
  \bibinfo{author}{\bibfnamefont{M.~F.} \bibnamefont{Schatz}},
  \bibinfo{journal}{Phys. Rev. Lett.} \textbf{\bibinfo{volume}{88}},
  \bibinfo{pages}{054501} (\bibinfo{year}{2002}).

\bibitem[{\citenamefont{Cross and Hohenberg}(1993)}]{ch93}
\bibinfo{author}{\bibfnamefont{M.~C.} \bibnamefont{Cross}} \bibnamefont{and}
  \bibinfo{author}{\bibfnamefont{P.~C.} \bibnamefont{Hohenberg}},
  \bibinfo{journal}{Rev. Mod. Phys.} \textbf{\bibinfo{volume}{65}},
  \bibinfo{pages}{851} (\bibinfo{year}{1993}).

\bibitem[{\citenamefont{Chandrasekhar}(1981)}]{c81}
\bibinfo{author}{\bibfnamefont{S.}~\bibnamefont{Chandrasekhar}},
  \emph{\bibinfo{title}{Hydrodynamic and Hydromagnetic Stability {\rm
  (International Series of Monographs on Physics (Oxford, England))}}}
  (\bibinfo{publisher}{Dover Publications}, \bibinfo{year}{1981}), ISBN
  \bibinfo{isbn}{978-0486640716}.

\bibitem[{\citenamefont{Morris et~al.}(1993)\citenamefont{Morris, Bodenschatz,
  Cannell, and Ahlers}}]{mbcdg93}
\bibinfo{author}{\bibfnamefont{S.~W.} \bibnamefont{Morris}},
  \bibinfo{author}{\bibfnamefont{E.}~\bibnamefont{Bodenschatz}},
  \bibinfo{author}{\bibfnamefont{D.~S.} \bibnamefont{Cannell}},
  \bibnamefont{and} \bibinfo{author}{\bibfnamefont{G.}~\bibnamefont{Ahlers}},
  \bibinfo{journal}{Phys. Rev. Lett.} \textbf{\bibinfo{volume}{71}},
  \bibinfo{pages}{2026} (\bibinfo{year}{1993}).

\bibitem[{\citenamefont{de~Bruyn et~al.}(1996)\citenamefont{de~Bruyn,
  Bodenschatz, Morris, Trainoff, Hu, Cannell, and Ahlers}}]{bbmthca96}
\bibinfo{author}{\bibfnamefont{J.~R.} \bibnamefont{de~Bruyn}},
  \bibinfo{author}{\bibfnamefont{E.}~\bibnamefont{Bodenschatz}},
  \bibinfo{author}{\bibfnamefont{S.~W.} \bibnamefont{Morris}},
  \bibinfo{author}{\bibfnamefont{S.~P.} \bibnamefont{Trainoff}},
  \bibinfo{author}{\bibfnamefont{Y.}~\bibnamefont{Hu}},
  \bibinfo{author}{\bibfnamefont{D.~S.} \bibnamefont{Cannell}},
  \bibnamefont{and} \bibinfo{author}{\bibfnamefont{G.}~\bibnamefont{Ahlers}},
  \bibinfo{journal}{Review of Scientific Instruments}
  \textbf{\bibinfo{volume}{67}}, \bibinfo{pages}{2043} (\bibinfo{year}{1996}).

\bibitem[{\citenamefont{Lam et~al.}(1992)\citenamefont{Lam, Lee, and
  Y.Suen}}]{lls92}
\bibinfo{author}{\bibfnamefont{L.}~\bibnamefont{Lam}},
  \bibinfo{author}{\bibfnamefont{S.-W.} \bibnamefont{Lee}}, \bibnamefont{and}
  \bibinfo{author}{\bibfnamefont{C.}~\bibnamefont{Y.Suen}},
  \bibinfo{journal}{IEEE Transactions on Pattern analysis and Machine
  Intelligence} \textbf{\bibinfo{volume}{14}}, \bibinfo{pages}{869}
  (\bibinfo{year}{1992}).

\bibitem[{\citenamefont{Walter et~al.}(2004)\citenamefont{Walter, Pesch, and
  Bodenschatz}}]{wpb04}
\bibinfo{author}{\bibfnamefont{T.}~\bibnamefont{Walter}},
  \bibinfo{author}{\bibfnamefont{W.}~\bibnamefont{Pesch}}, \bibnamefont{and}
  \bibinfo{author}{\bibfnamefont{E.}~\bibnamefont{Bodenschatz}},
  \bibinfo{journal}{Chaos: An Interdisciplinary Journal of Nonlinear Science}
  \textbf{\bibinfo{volume}{14}}, \bibinfo{pages}{933} (\bibinfo{year}{2004}),
  \urlprefix\url{http://link.aip.org/link/?CHA/14/933/1}.

\bibitem[{\citenamefont{Busse}(1967)}]{b67}
\bibinfo{author}{\bibfnamefont{F.~H.} \bibnamefont{Busse}},
  \bibinfo{journal}{Journal of Fluid Mechanics} \textbf{\bibinfo{volume}{30}},
  \bibinfo{pages}{625} (\bibinfo{year}{1967}).

\bibitem[{piv()}]{piv}
\bibinfo{howpublished}{\url{http://www.physics.uci.edu/~foams}}.

\bibitem[{\citenamefont{Huepe et~al.}(2004)\citenamefont{Huepe, Riecke,
  Daniels, and Bodenschatz}}]{hrdb04}
\bibinfo{author}{\bibfnamefont{C.}~\bibnamefont{Huepe}},
  \bibinfo{author}{\bibfnamefont{H.}~\bibnamefont{Riecke}},
  \bibinfo{author}{\bibfnamefont{K.~E.} \bibnamefont{Daniels}},
  \bibnamefont{and}
  \bibinfo{author}{\bibfnamefont{E.}~\bibnamefont{Bodenschatz}},
  \bibinfo{journal}{Chaos: An Interdisciplinary Journal of Nonlinear Science}
  \textbf{\bibinfo{volume}{14}}, \bibinfo{pages}{864} (\bibinfo{year}{2004}),
  \urlprefix\url{http://link.aip.org/link/?CHA/14/864/1}.

\bibitem[{\citenamefont{Riecke and Madruga}(2006)}]{rm06}
\bibinfo{author}{\bibfnamefont{H.}~\bibnamefont{Riecke}} \bibnamefont{and}
  \bibinfo{author}{\bibfnamefont{S.}~\bibnamefont{Madruga}},
  \bibinfo{journal}{Chaos: An Interdisciplinary Journal of Nonlinear Science}
  \textbf{\bibinfo{volume}{16}}, \bibinfo{eid}{013125}
  (pages~\bibinfo{numpages}{11}) (\bibinfo{year}{2006}),
  \urlprefix\url{http://link.aip.org/link/?CHA/16/013125/1}.

\bibitem[{\citenamefont{Gameiro et~al.}(2004)\citenamefont{Gameiro, Mischaikow,
  and Kalies}}]{gmk04}
\bibinfo{author}{\bibfnamefont{M.}~\bibnamefont{Gameiro}},
  \bibinfo{author}{\bibfnamefont{K.}~\bibnamefont{Mischaikow}},
  \bibnamefont{and} \bibinfo{author}{\bibfnamefont{W.}~\bibnamefont{Kalies}},
  \bibinfo{journal}{Phys. Rev. E} \textbf{\bibinfo{volume}{70}},
  \bibinfo{pages}{035203} (\bibinfo{year}{2004}).

\end{thebibliography}
\end{document}